\documentclass[prl,twocolumn,nofootinbib,floatfix,10pt]{revtex4-2}

\usepackage{amsmath}
\usepackage{amssymb}
\usepackage{wasysym}
\usepackage{graphicx}
\usepackage{color,soul}
\usepackage{physics}
\usepackage{siunitx}
\usepackage{dsfont}
\usepackage{float}
\usepackage[english]{babel}
\usepackage{blindtext}
\usepackage[english,nomargin,inline,marginclue,draft]{fixme}
\pdfpageheight\paperheight
\pdfpagewidth\paperwidth

\usepackage[colorlinks,linkcolor=blue,anchorcolor=blue,citecolor=blue,urlcolor=blue]{hyperref}

\fxusetheme{colorsig}
\FXRegisterAuthor{cg}{acg}{CG}  
\FXRegisterAuthor{th}{ath}{\color{blue}TH}  
\FXRegisterAuthor{ib}{aib}{\color{red}IB} 
\FXRegisterAuthor{sh}{ash}{\color{cyan}SH} 
\FXRegisterAuthor{db}{adb}{\color{green}DB} 
\FXRegisterAuthor{ps}{aps}{PS}
\makeatletter
\renewcommand*\FXLayoutInline[3]{%
  {\@fxuseface{inline}\ignorespaces{\color{fx#1}[#3: #2]}}}
\makeatother

\long\def\symbolfootnote[#1]#2{\begingroup%
\def\thefootnote{\fnsymbol{footnote}}\footnotetext[#1]{#2}\endgroup}

\def\nobreakbefore{%
  \relax\ifvmode\else
    \ifhmode
      \ifdim\lastskip > 0pt\relax
        \unskip\nobreakspace
      \else 
        \nobreakspace
      \fi
    \fi
  \fi
}
\let\oldcite\cite
\renewcommand\cite{\nobreakbefore\oldcite}

\begin{document}

\title{Microwave control of collective quantum jump statistics of a dissipative Rydberg gas}

\author{Zong-Kai Liu$^{1,2,\dagger}$}

\author{Kong-Hao Sun$^{1,2,\dagger}$}

\author{Albert Cabot$^{3}$}

\author{Federico Carollo$^{3}$}

\author{Jun Zhang$^{1,2}$}

\author{Zheng-Yuan Zhang$^{1,2}$}

\author{Li-Hua Zhang$^{1,2}$}
\author{Bang Liu$^{1,2}$}

\author{Tian-Yu Han$^{1,2}$}
\author{Qing Li$^{1,2}$}

\author{Yu Ma$^{1,2}$}
\author{Han-Chao Chen$^{1,2}$}

\author{Igor Lesanovsky$^{3,4}$}

\author{Dong-Sheng Ding$^{1,2,\textcolor{blue}{\star}}$}

\author{Bao-Sen Shi$^{1,2}$}

\affiliation{$^1$Key Laboratory of Quantum Information, University of Science and Technology of China, Hefei, Anhui 230026, China.}
\affiliation{$^2$Synergetic Innovation Center of Quantum Information and Quantum Physics, University of Science and Technology of China, Hefei, Anhui 230026, China.}
\affiliation{$^3$Institut für Theoretische Physik, Eberhard Karls Universität Tübingen, Auf der Morgenstelle 14, 72076 Tübingen, Germany}
\affiliation{$^4$School of Physics and Astronomy and Centre for the Mathematics and Theoretical Physics of Quantum
Non-Equilibrium Systems, The University of Nottingham, Nottingham, NG7 2RD, United Kingdom}

\date{\today}

\symbolfootnote[1]{dds@ustc.edu.cn}
\symbolfootnote[2]{Z.K.L and K.H.S contribute equally to this work.}

\begin{abstract}
Quantum many-body systems near phase transitions respond collectively to externally applied perturbations. We explore this phenomenon in a laser-driven dissipative Rydberg gas that is tuned to a bistable regime. Here two metastable phases coexist, which feature a low and high density of Rydberg atoms, respectively. The ensuing collective dynamics, which we monitor in situ, is characterized by stochastic collective jumps between these two macroscopically distinct many-body phases. We show that the statistics of these jumps can be controlled using a dual-tone microwave field. In particular, we find that the distribution of jump times develops peaks corresponding to subharmonics of the relative microwave detuning. Our study demonstrates the control of collective statistical properties of dissipative quantum many-body systems without the necessity of fine-tuning or of ultra cold temperatures. Such robust phenomena may find technological applications in quantum sensing and metrology.  
\end{abstract}

\maketitle

Open many-body quantum systems in which dissipative processes and coherent interactions compete may display emergent behavior. This manifests in novel timescales and dynamical regimes, which are not simply predictable from the knowledge of the underlying microscopic physics alone. For example, symmetries of the microscopic equations of motion can be spontaneously broken and as a consequence non-ergodic dynamical behavior occurs. Further to that, the entire system responds collectively to externally applied perturbations. Not only can such mechanism be exploited in practical applications, such as collectively enhanced sensing devices \cite{aldridge2005,Wade2018,raghunandan2018}, but it  may also underlie the physics of learning, as demonstrated in the case of pattern retrieval dynamics in the Hopfield neural network model \cite{hopfield1982}.

In recent years Rydberg gases have become a widely employed system for the investigation of many-body physics. The Rydberg platform features strong long-range interactions \cite{saffman2010quantum,adams2019rydberg, browaeys2020many} together with dissipation channels, which enabled the study of nonequilibrium phase transitions \cite{ lee2012collective,carr2013nonequilibrium,schempp2014full,marcuzzi2014universal,lesanovsky2014out, urvoy2015strongly,ding2022enhanced}, the observation of signatures of self-organization \cite{ding2019Phase,Signatures2020Helmrich,Wintermantel2020Cellular, klocke2021hydrodynamic} as well as of ergodicity breaking and synchronization \cite{gambetta2019, Wadenpfuhl2023Synchronization, ding2023ergodicity, wu2023}. 

Here we demonstrate an experiment conducted in the metastable (or bistable) \cite{lee2012collective,marcuzzi2014universal,wade2017real,Bistability2017Letscher} regime of a dissipative Rydberg gas, where two phases – one with low and one with high Rydberg density – coexist. Dynamically this coexistence manifests in the random switching of the Rydberg density, with exponentially distributed switching times. We monitor this macroscopic effect continuously and non-destructively via an electromagnetically induced transparency (EIT) detection method. We show that by applying a periodic external perturbation, via a dual-tone microwave (MW) electric field, the collective dynamical behavior of the Rydberg gas can be dramatically altered. This manifests directly in the statistics of the switching times, which ceases to be continuous and  develops instead discrete peaks at multiples of the period associated with the relative detuning of the MW drive.

\begin{figure}[t]
\includegraphics[width=1\columnwidth]{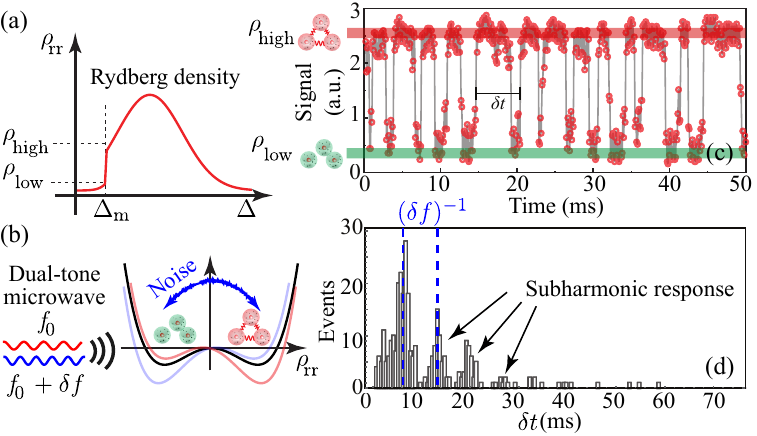}\caption{\textbf{Collective quantum jumps.} (a) Sketch of the stationary density in an ensemble of laser-driven Rydberg atoms as a function of the detuning $\Delta$ of the Rydberg excitation laser. The presence of non-linear effects can lead to the emergence of a bistable regime. This manifests as a sudden change in the Rydberg density occurring at a detuning $\Delta_m$. This indicates that there are two (metastable) states or phases, one with a high density of Rydberg excitations ($\rho_{\mathrm{high}}$) and one with a low density ($\rho_{\mathrm{low}}$). (b) The central features of the physics in this bistable regime are effectively captured by a double-well model. The two wells correspond to the two phases and the presence of noise induces collective jumps between them [see panel (c)]. Driving the Rydberg manifold via dual-tone MW fields, with frequencies $f_0$ and $f_0 + \delta f$, periodically modulates the potential landscape (see different curves). (c) Collective jumps between the two phases appear in the time-resolved transmission signal, here in the presence of the MW field (a.u. stands for arbitrary units, as the transmission signal is obtained through the voltage of the detector). (d) Histogram of the time intervals $\delta t$ between consecutive upwards jumps (from low to high Rydberg density) for the data in panel (c). In the bistable regime, the system responds collectively to the MW. Here, the histogram develops characteristic peaks corresponding to subharmonics of the frequency difference $\delta f$ between the dual-tone MW field components. 
}
\label{setup}
\end{figure}

Before going to the detailed description of our experimental findings, we illustrate the central idea of our work in Fig.~\ref{setup}. In our setup we excite Rydberg states from hot atomic vapour, where the atomic density is sufficiently large such that atoms in Rydberg states strongly interact. As shown in a number of previous works \cite{weller2016charge, de2016intrinsic,Bistability2017Letscher,Interplay2019Weller}, this induces non-linear and bistable behavior. The latter manifests for instance in a sudden jump from low density $\rho_{\mathrm{low}}$ to high density $\rho_{\mathrm{high}}$ of Rydberg atoms, $\rho_{rr}$, when one of the system parameter is tuned. In Fig.~\ref{setup}(a) this is sketched for a situation in which the detuning $\Delta$ of the excitation laser is varied. This sudden change can be interpreted \cite{marcuzzi2014universal} by evoking an analogy with equilibrium thermodynamics: one may assume that the stationary state in the bistable regime is governed by an effective potential, which exhibits a double-well shape, as sketched in Fig.~\ref{setup}(b). The two minima correspond to two phases. Close to the transition point, i.e., near $\Delta_m$ shown in Fig.~\ref{setup}(a), the system performs collective jumps, switching between the two phases with macroscopically distinct Rydberg densities (see transmission signal in Fig.~\ref{setup}(c) which shows data obtained using our EIT detection method). This qualitative picture can be put on a solid theoretical footing using the theory of metastablity \cite{macieszczak2016,macieszczak2021}, and in the Supplemental Material we provide more details on this aspect. The switching is stochastic, i.e., it is driven by noise, and the characteristic time $\delta t$ between consecutive jumps from the low-density towards the high-density phase follows an exponential distribution (see Supplemental Material). As shown in Fig.~\ref{setup}(d), the application of a dual-tone MW field, with relative frequency difference $\delta f$ [see panel (b)] leads to a non-continuous distribution for the  times $\delta t$. In particular, the resulting distribution exhibits peaks at times that are multiples of $1/\delta f$.

\begin{figure*}
\centering
\includegraphics[width=1.9\columnwidth]{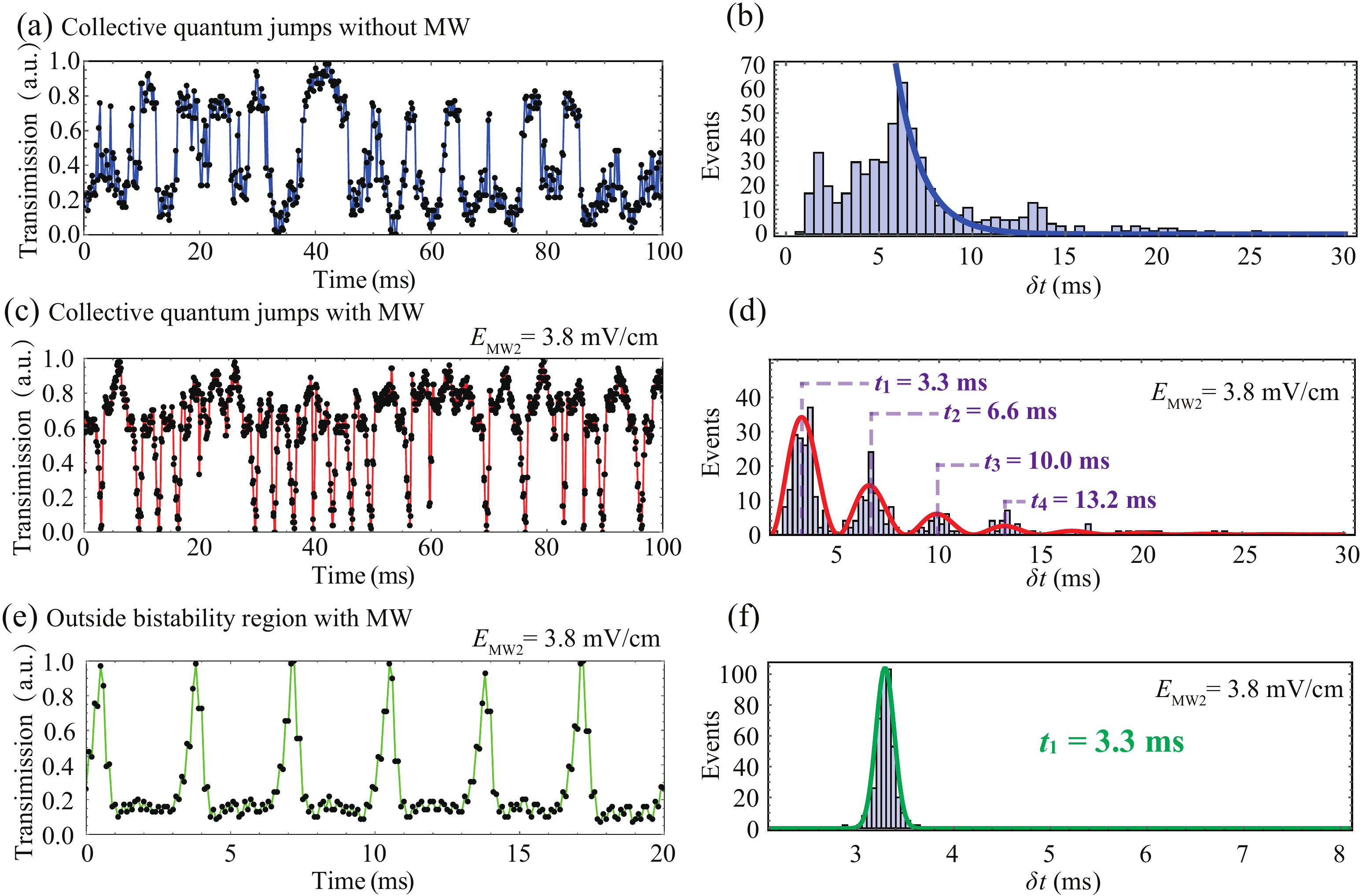}\\
\caption{\textbf{Controlling the collective quantum jump statistics.} (a) Transmission signal showing collective jumps in the absence of MW driving in the bistable region [cf.~Fig.~\ref{setup}(a,b,c)]. (b) The corresponding histogram of the time interval $\delta t$ between consecutive upwards jumps from low to high density [see Fig. \ref{setup}(c)] displays exponential behavior, given by the  fit $g_1(\delta t)\propto\times\exp(-0.7\, \delta t/\mathrm{ms} )$  (solid line). (c) Transmission signal in the presence of MW driving. (d) The histogram for the case in (c) displays peaks occurring at multiples of the modulation period (see labels). The solid line corresponds to the fit $g_2(\delta t) \propto \exp(-0.26\,\delta t /\mathrm{ms})[\sin(1.88\delta t/\mathrm{ms}+\pi/2)+1]$. (e) Transmission signal far from the bistable regime and in the presence of MW driving. (f) Histogram of the time intervals between peaks in (e). The solid line corresponds to the fit $g_3(\delta t) \propto \exp(-56\,[\delta t/\text{ms}-3.3]^{2})$. All data are obtained for the same detuning $\Delta_c$ and $\Delta_p$. In panels (a,b) there is no MW. In panels (c-f) the dual-tone MW field of amplitude $E_{\text{MW2}} = 3.8$ mV/cm is applied. In panels (e,f) the Rabi frequency of the probe light $\Omega_p$ is so low that there is no bistability.}\label{information}
\end{figure*}

\textbf{Experimental setup.} In our experiment rubidium-85 atoms are placed into a $10$ cm long heated glass cell and the temperature is stabilized to $45.0^\circ \mathrm{C}$ (atomic density $9.19\times 10^{10}  \text{ } \mathrm{cm}^{-3}$). Atoms are excited from the ground state ($5\mathrm{S}_{1/2}, F=2$) to the Rydberg state ($51\mathrm{D}_{3/2}$) via a two-photon process. A 795 nm probe light (with $1/e^2$-waist radius of approximately 500 $\mathrm{\mu m}$  and peak value of Rabi frequency $\Omega_p/2\pi$ $\sim1 \text{ } \mathrm{MHz}$) excites atoms from the ground state to the intermediate level ($5P_{1/2}, F=3$) and a further coupling laser with wavelength 480 nm (with $1/e^2$-waist radius of approximately 200 $\mathrm{\mu m}$ and peak value of Rabi frequency $\Omega_c/2\pi$ $\sim 10 \text{ } \mathrm{MHz}$) takes atoms to the Rydberg state. This is an EIT configuration, where the coupling light controls the absorption of the probe light. 
To observe the EIT spectrum, the detuning of the coupling light $\Delta_c$ is scanned while the probe light detuning $\Delta_p$ is resonant to the transition  $5\mathrm{S}_{1/2} - 5P_{1/2}$. The Rabi frequency of the probe light and its detuning are both fixed. After setting the detuning of the coupling light to a specific value $\Delta_c$, the time evolution of the transmission signal is obtained, yielding curves as the one depicted in Fig.~\ref{information}(a).  In this way, we can study the collective response of the Rydberg gas through the time-dependent transmission. 

Besides the probe and coupling beams, there is a dual-tone MW field applied to Rydberg atoms by a MW generator and a horn close to the rubidium glass cell. One MW field is resonant with the Rydberg transition $51\mathrm{D}_{3/2}-52\mathrm{P}_{1/2}$. Here, $51\mathrm{D}_{3/2}$ is the Rydberg state in EIT configuration. A second MW field is detuned from this resonance by an amount $\delta f$. Specifically, this dual-tone MW field has the form  $E_{\text{MW1}}\sin(2\pi f_{0}t)+ E_{\text{MW2}}\,\sin(2\pi[f_{0}+\delta f]t)$ where $f_{0}$ = $16.67$ GHz is near resonant with the $51D_{3/2} - 52P_{1/2}$ transition and $\delta f = 300$ Hz represents the relative MW field detuning. The amplitude of the first MW field is fixed at $E_{\text{MW1}}=3.8 \mathrm{mV/cm}$ and that of the second MW field can be attenuated during the experiment. The effective Rabi frequency of the dual-tone MW field is then modulated according to $\Omega_{\text{MW}}(t)=\Omega_{\text{MW1}}+\Omega_{\text{MW2}} \exp (-i 2\pi \delta f t)$, where $\Omega_{\text{MW1(2)}}$ corresponds to the Rabi frequency of the MW field resonant (detuned by $\delta f$) to the Rydberg states. The MW frequency resonant to the transition $51D_{3/2}  - 52P_{1/2}$ is calculated via the method reported in Ref. \cite{vsibalic2017arc}. The amplitude of the MW in the center of the rubidium cell is calibrated by peak splitting of the EIT spectrum according to the Autler-Townes split \cite{sedlacek2012microwave} (for more details see Supplemental Material).

\begin{figure}
\centering
\includegraphics[width=1\columnwidth]{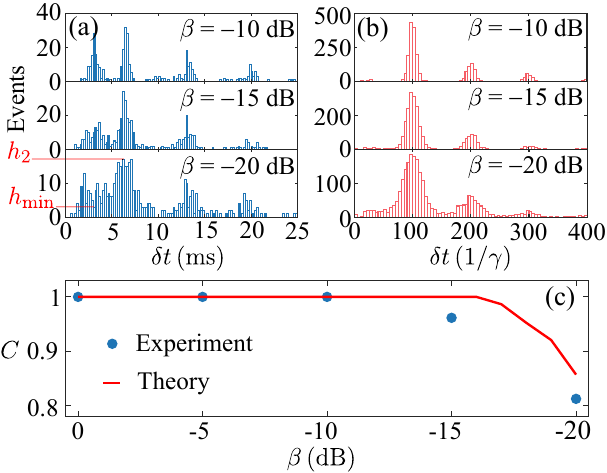}
\caption{\textbf{Jump time distributions as a function of MW intensity.} (a) Experimentally measured distributions of $\delta t$ for different values of $\beta$, from $-10$ dB to $-20$ dB,  and  $\delta f =300$ Hz.  (b) Distributions of $\delta t$ obtained with our theoretical model for values of $\beta$ as in (a) and $\delta f= 0.01 \gamma$. Here $\gamma$ is the lifetime of the Rydberg state in our theoretical model. Details are given in the Supplemental Material. (c) Behavior of the contrast $C=(h_2-h_{\mathrm{min}})/h_2$ as a function of the parameter $\beta$. Here, $h_2$ is the hight of the second peak and $h_{\text{min}}$ is the minimum height of the bins between the first and the second peak (see bottom of panel (a) for an example).
We display results from both experimental (bullet) and theoretical (solid line) data.}\label{num1}
\end{figure}

\textbf{Statistics of collective quantum jumps.} As discussed previously and sketched in Fig.~\ref{setup}, stochastic collective jumps manifest in the transmission signal. To establish a benchmark, we show this phenomenon in Fig.~\ref{information}(a) in the absence of the MW field. We choose Rabi frequency and detuning of the probe and coupling light such that the system is located in a bistable parameter region. The physical process underlying the collective jumps was discussed in Ref. \cite{lee2012collective}. The jump process from the low to the high density phase (upward jump) can be understood as being induced by a fluctuation of the Rydberg population which triggers an avalanche of Rydberg excitations. Specifically, when one atom happens to be excited from the ground to the Rydberg state by a fluctuation, it shifts the Rydberg energy level of the surrounding atoms \cite{schempp2014full,urvoy2015strongly,guttierez2017,Signatures2020Helmrich} and makes it precisely resonant with the detuned two-photon excitation. This starts a facilitation or avalanche process which leads to a sharp increase of the Rydberg population. As argued in Ref. \cite{lee2012collective} also the downward jump (from high to low density) is collective: when the system resides in the high-density phase and there are only very few Rydberg atoms decay over some period of time, the non-unitary no-decay  evolution governs the dynamics and enhances the weight of configurations hosting only few Rydberg excitations. This brings the system to the low-density phase. 

On a more formal level, the intermittent dynamics originating from the collective jumps can be understood from the theory of metastability \cite{macieszczak2016,macieszczak2021}. Here the low- and the high-density phase correspond to two metastable states that are approximate steady-states of the dynamical generator that governs the time-evolution of the dissipative Rydberg gas. These two states are connected via an effective classical stochastic dynamics, that entails collective jumps between them. This predicts that the time interval $\delta t$ between two consecutive upwards jumps follows an exponential distribution for sufficiently large $\delta t$ (see Supplemental Material). As shown in Figs.~\ref{information}(a,b) this is indeed confirmed by our experiment. To produce histograms as shown in Fig.~\ref{information}(b) we collect $20$ sets of transmission signals, one of which is shown in Fig.~\ref{information}(a), and count how often a time interval of length $\delta t$ occurs between two consecutive upward jumps (from the low- to the high-density phase). The details of the procedure for generating the histograms from the transmission signals are given in the Supplemental Material. 
Note, that at small $\delta t$, there is hole in the histogram. This is a consequence of ``anti-bunching" in two-level systems: after an upward jump, a downward jump has to occur first, before the next upward jump can take place. This limits the smallest value of $\delta t$, that is observable. 

In the second and third row of Fig.~\ref{information} we show data obtained in the presence of the dual-tone MW field. We distinguish here the case in which the system is initially in the bistable region [Figs.~\ref{information}(c,d)] from the case in which the system is initially far from it [Figs.~\ref{information}(e,f)]. For a bistable system the application of the MW leads to a drastic change of the distribution of the jump times, as shown in  Fig.~\ref{information}(d). The histogram breaks up into multiple  disconnected peaks with an exponentially decaying envelope. This is indeed a collective response to the dual-tone MW, which can be seen by making a comparison with the case in which the system is far from a bistable regime. Here, the application of the MW with oscillating Rabi frequency $\Omega_{\text{MW}}$, merely modulates the transmission signal monochromatically, see Fig.~\ref{information}(e).  
The associated histogram, shown in Fig.~\ref{information}(f), displays a single peak at $\delta t = t_1 \approx 3.3 \mathrm{ms}$, which corresponds to the relative detuning of dual-tone MW field: $\delta f = 300 \text{ } \mathrm{Hz}$. The bistable system, however, responds not only at that frequency, but also features collective subharmonic responses: a second, a third, and a forth peak appear in Fig.~\ref{information}(d), labeled as  $t_2  = 6.6\text{ } \mathrm{ ms}$,  $t_3  = 10.0\text{ } \mathrm{ms}$, and $t_4  = 13.2\text{ } \mathrm{ms}$.

\textbf{Tunable response.} To  characterize  the observed subharmonic response, we vary the amplitude of the second MW field $E_{\text{MW2}}$, as quantified by the parameter $\beta=20\log_{10}\left(E_{\text{MW2}}/E_{\text{MW1}}\right)$, and the frequency difference $\delta f$ (see Supplemental  Material). As shown in Fig.~\ref{num1}(a), upon decreasing $E_{\text{MW2}}$ the subharmonic peaks decrease in height and broaden, while remaining in their position. As apparent in the bottom panel of Fig.~\ref{num1}(a), the distribution approaches a continuous one for sufficiently weak driving $E_{\text{MW2}}$. 
To understand the origin of the subharmonic response, we consider a simple mean-field model described by a Lindblad master equation~\cite{Wade2018}. At the mean-field level, the presence of many-body interactions between Rydberg atoms results in a non-linear shift of the detuning which depends on the Rydberg state density. In addition, we include a noise term in the detuning, encoding thermal fluctuations (see discussion below). The latter term is crucial for the emergence of the subharmonic response (for details on the theoretical model see Supplemental Material). The distribution of $\delta t$ calculated with our model is illustrated in Fig.~\ref{num1}(b), as a function of the amplitude of the second MW field, $E_{\text{MW2}}$. The theoretical results show good qualitative agreement with our experiment. 

The change of the jump time distribution upon decreasing the driving intensity $E_{\text{MW2}}$ can be understood in terms of the double-well model discussed at the beginning [see Fig.~\ref{setup}(b)]. The driving modulates the ``potential landscape" periodically in a way that gives rise to an optimal time within the period, at which collective jumps are favored. When the modulation is strong in comparison to the noise this leads to narrow peaks in the distribution [cf.~Fig.~\ref{information}(c)]. However, as the driving weakens, the modulation of the potential is less effective and noise becomes the main driver for the collective jumps. In this case, there is no preferred time for their occurrence and the distribution broadens. 

To characterize the broadening of the peaks, we define the quantity $C=\left(h_2-h_\text{min}\right)/h_2$, which measures the contrast between the height of the second peak, $h_2$, and the minimum height of the bins between the first and the second peak, $h_{\text{min}}$. As illustrated in Fig.~\ref{num1}(c), when the amplitude of the second MW field is decreased below a threshold value, the contrast becomes smaller than one, indicating coalescence of the first and the second peak. This behavior is also captured by our theoretical model.

\textbf{Discussion.} We have demonstrated that driving a bistable Rydberg-atom system, via a dual-tone MW field, can change the distribution of the times between collective jumps. The latter passes from a late time exponential distribution to one with several peaks corresponding to subharmonics of the relative MW detuning. 
Details of the distribution can be  controlled by varying the amplitude and the detuning of the dual-tone MW fields. 

The emergence of a subharmonic response is crucially rooted in the presence of noise. This is primarily induced by thermal fluctuations of the number of particles in the excitation area, where the probe light and the coupling light overlap. These fluctuations effectively induce a noisy (mean-field) shift on the detuning. 
For increasing temperatures, the number of atoms $N$ in the excitation area increases. As a consequence, the relative fluctuation $\delta N/N$ gradually decreases ($\delta N$ is the fluctuation of atom number), resulting in a vanishing detuning noise and in the disappearance of the subharmonic response. The above discussion aligns with experiments conducted at higher temperatures.
Interestingly, the observed phenomenology  
is not limited to Rydberg-atom systems but 
can also occur, for instance, in bistable neuronal systems \cite{Rose1967Nerve, longtin1994bistability}.

\textit{Note.} While preparing this manuscript we became aware of a related work (Ref.~\cite{wu2024nonlinearityenhanced}), where the collective response of a dissipative Rydberg gas is employed for sensing MW fields.

\begin{acknowledgements}
\textbf{Acknowledgments.} D-S.D thanks for discussions with professor Jun Ye from JILA, and appreciates instructive comments from Dr. Daniel Malz, Prof. Charles S. Adams and Prof. Wei Yi. We acknowledge funding from the National Key R\&D Program of China (Grant No. 2022YFA1404002), the National Natural Science Foundation of China (Grant Nos. U20A20218, 61525504, and 61435011), the Anhui Initiative in Quantum Information Technologies (Grant No. AHY020200), the major science and technology projects in Anhui Province (Grant No. 202203a13010001). AC and IL acknowledge funding from the Deutsche Forschungsgemeinschaft (DFG, German Research Foundation) through the Research Units FOR 5413/1, Grant No. 465199066, and the Walter Benjamin programme, Grant No. 519847240. IL also received funding from the European Union’s Horizon Europe research and innovation program under Grant Agreement No. 101046968 (BRISQ). FC~is indebted to the Baden-W\"urttemberg Stiftung for the financial support of this research project by the Eliteprogramme for Postdocs.
\end{acknowledgements}

\bibliography{ref}

\end{document}


\title{Supplemental material for Microwave control of collective quantum jump statistics of a dissipative Rydberg gas}

\author{Zong-Kai Liu$^{1,2,\dagger}$}

\author{Kong-Hao Sun$^{1,2,\dagger}$}

\author{Albert Cabot$^{3}$}

\author{Federico Carollo$^{3}$}

\author{Jun Zhang$^{1,2}$}

\author{Zheng-Yuan Zhang$^{1,2}$}

\author{Li-Hua Zhang$^{1,2}$}
\author{Bang Liu$^{1,2}$}

\author{Tian-Yu Han$^{1,2}$}
\author{Qing Li$^{1,2}$}

\author{Yu Ma$^{1,2}$}
\author{Han-Chao Chen$^{1,2}$}

\author{Igor Lesanovsky$^{3,4}$}

\author{Dong-Sheng Ding$^{1,2,\textcolor{blue}{\star}}$}
\author{Bao-Sen Shi$^{1,2}$}

\affiliation{$^1$Key Laboratory of Quantum Information, University of Science and Technology of China, Hefei, Anhui 230026, China.}
\affiliation{$^2$Synergetic Innovation Center of Quantum Information and Quantum Physics, University of Science and Technology of China, Hefei, Anhui 230026, China.}
\affiliation{$^3$Institut für Theoretische Physik, Eberhard Karls Universität Tübingen, Auf der Morgenstelle 14, 72076 Tübingen, Germany}
\affiliation{$^4$School of Physics and Astronomy and Centre for the Mathematics and Theoretical Physics of Quantum
Non-Equilibrium Systems, The University of Nottingham, Nottingham, NG7 2RD, United Kingdom}

\date{\today}

\symbolfootnote[1]{dds@ustc.edu.cn}
\symbolfootnote[2]{Z.K.L and K.H.S contribute equally to this work.}

\renewcommand{\theequation}{S\arabic{equation}}
\renewcommand{\thefigure}{S\arabic{figure}}

\maketitle
\section{Details on computing the histogram for collective jump times}
\begin{figure*}
\centering
\includegraphics[width=\columnwidth]{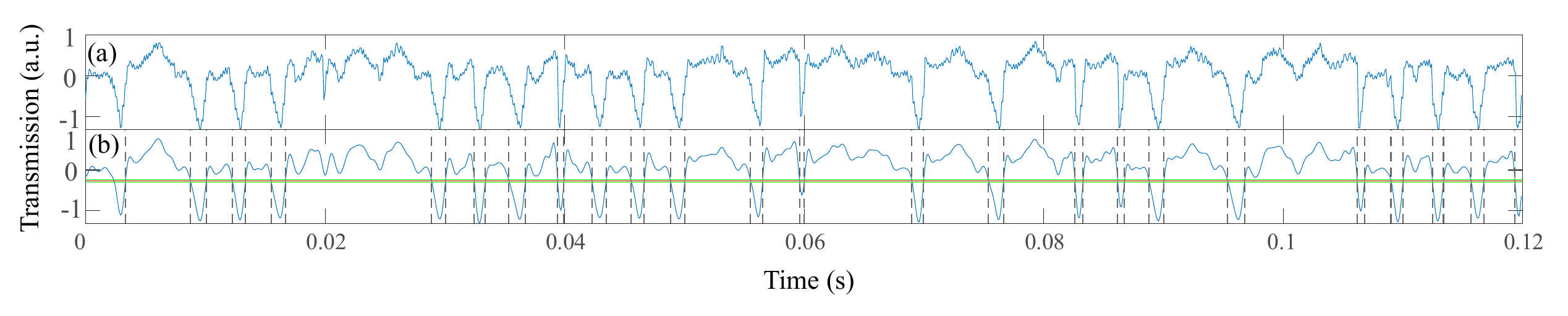}\\
\caption{(a) Unprocessed transmission signal. Note that the $y$-axis is in arbitrarily chosen units. (b) The signal is run through a low-pass filter in order to remove high frequency noise. To determine the times at which collective quantum jumps occur we impose a threshold value $\mu$ (red line) that separates the low density and high density phase. Due to the discreteness of the data we define in practice a threshold interval ($\mu-\alpha$, $\mu+\alpha$) delimited by the two green lines (here $\alpha = 3.2 \times 10^{-2}$). The jump times (indicated by the vertical lines) are estimated by identifying the times at which the transmission signal passes through the threshold region.}
\label{noMW_xline}
\end{figure*}

In this section we provide more details about how to obtain the data for the histograms (as shown in Figs.~2(b, d, f), Fig.~3(a) of the main text) from experimental trajectories (Figs.~2(a, c, e) of the main text). First, we apply a low-pass filter and a linear interpolation to the trajectory data [see Fig.~\ref{noMW_xline}(a)]. This yields a smoothed curve, shown in panel (b). The reason for this interpolation is that the resolution of the oscilloscope in our experiment is too low in order to yield a dense sampling of the transmission signal. Afterwards, we select a suitable transmission threshold value, $\mu$, indicated by the red line in Fig.~\ref{noMW_xline}(b), which discriminates between the high and low density phase. However, for processing our discrete data we need to define a finite transition interval ($\mu-\alpha$, $\mu+\alpha$), indicated by two green horizontal lines. When the transmission curve passes through the threshold region a collective quantum jump is identified (see dashed vertical lines). Selecting only upward jumps and calculating the time difference of consecutive ones yields the $\delta t$-values for which the histogram is shown.

\section{Theoretical Model}
To understand the sub-harmonic response, a model based on the Lindblad master equation is proposed. The three-level system is described by the Lindblad master equation:
\begin{align}
    \frac{\partial \rho(t)}{\partial t}=-i[H(t), \rho(t)]+\sum_{i=1}^2\left(L_i \rho(t) L_i^{\dagger}-\frac{1}{2}\left\{L_i^{\dagger} L_i, \rho(t)\right\}\right)\, .
\end{align}
In the rotating frame, the time-dependent Hamiltonian $H(t)$ has
the form
\begin{align}
    H(t)=\left(\begin{array}{ccc}
0 & \Omega / 2 & 0 \\
\Omega^* / 2 & -\Delta_\text{eff1} & \Omega_s(t) / 2 \\
0 & \Omega_s^*(t) / 2 & -\Delta_\text{eff2}
\end{array}\right),
\end{align}
in the basis $\{|g\rangle,|r\rangle,|s\rangle\}$,
where $|g\rangle$ is the ground state and $\{|r\rangle,|s\rangle\}$ are the Rydberg states. The Rabi frequency of the dual-tone MW is $\Omega_s(t)=\Omega_{\text{MW1}}+\Omega_{\text{MW2}} \exp (-i2\pi\delta f t)$, where $\Omega_{\text{MW1(2)}}$ corresponds to the Rabi frequency of the MW resonant (detuned by $\delta f$) to the Rydberg transition. The effective detunings $\Delta_\text{eff1}=\Delta-V_1 \rho_{r r}-V_2 \rho_{s s}$ and $\Delta_\text{eff2}=\Delta-V_3 \rho_{r r}-V_4 \rho_{s s}$ originate from many-body interaction under a mean-field approximation or plasma induced stark shift on Rydberg levels. The jump operators $L_1=\sqrt{\gamma_r}|g\rangle\langle r|$ and $L_2=\sqrt{\gamma_s}|g\rangle\langle s|$ account for the decay channels
$|r\rangle \rightarrow|g\rangle$ and $|s\rangle \rightarrow|g\rangle$ respectively. Besides considering interactions within a mean-field approximation, we additionally include a noise term in the detuning $\Delta + dW$, with $dW$ being a Gaussian white noise.

In Fig.~3(d-g) in the main text, we show the sub-harmonic response in the jump interval distribution as well as the contrast between the first and second peaks calculated by the numerical model, as the function of driving intensity. The parameters are: $\gamma_r=\gamma_s\equiv\gamma$, $\Omega=\gamma$, $V_1=-10\gamma$, $V_2=-5\gamma$, $V_3=-30\gamma$, $V_4=-15\gamma$, $\delta f=0.01\gamma$, $\Omega_{\text{MW1}}=3\gamma$, $\Omega_{\text{MW2}} = 0.949\gamma, 0.533\gamma, 0.3\gamma$ and $\Delta=-4.15\gamma, -3.95\gamma, -3.87\gamma$ in panels (d-f) respectively. The standard deviation of Gaussian noise $dW$ is $0.2\gamma$.

Here the white noise $dW$ is generated by creating a signal consisting of $\gamma T/10$ data points ($T=1/\delta f$), where each point is sampled from a normal distribution with a mean of zero and a standard deviation of $0.2\gamma$. The resulting signal is then interpolated to obtain a continuous function over the time interval $[0, 100T]$. In this model, we define the detuning that maximizes the sub-harmonic response as the optimum point. Specifically, we take the detuning that maximizes the count within the jump interval $[1.85T, 2.15T]$ as the optimum point in numerical calculation. To get the jump interval distribution, we consider $32$ evolution trajectories of
Rydberg states population $n_\text{R}=\rho_{22}+\rho_{33}$ for statistical analysis, each with an evolution time of $10000/\gamma$, and the bin width of histograms are set to $T/20$ in Fig.~3 for both experiment and theory.

\begin{figure*}
\centering
\includegraphics[width=\columnwidth]{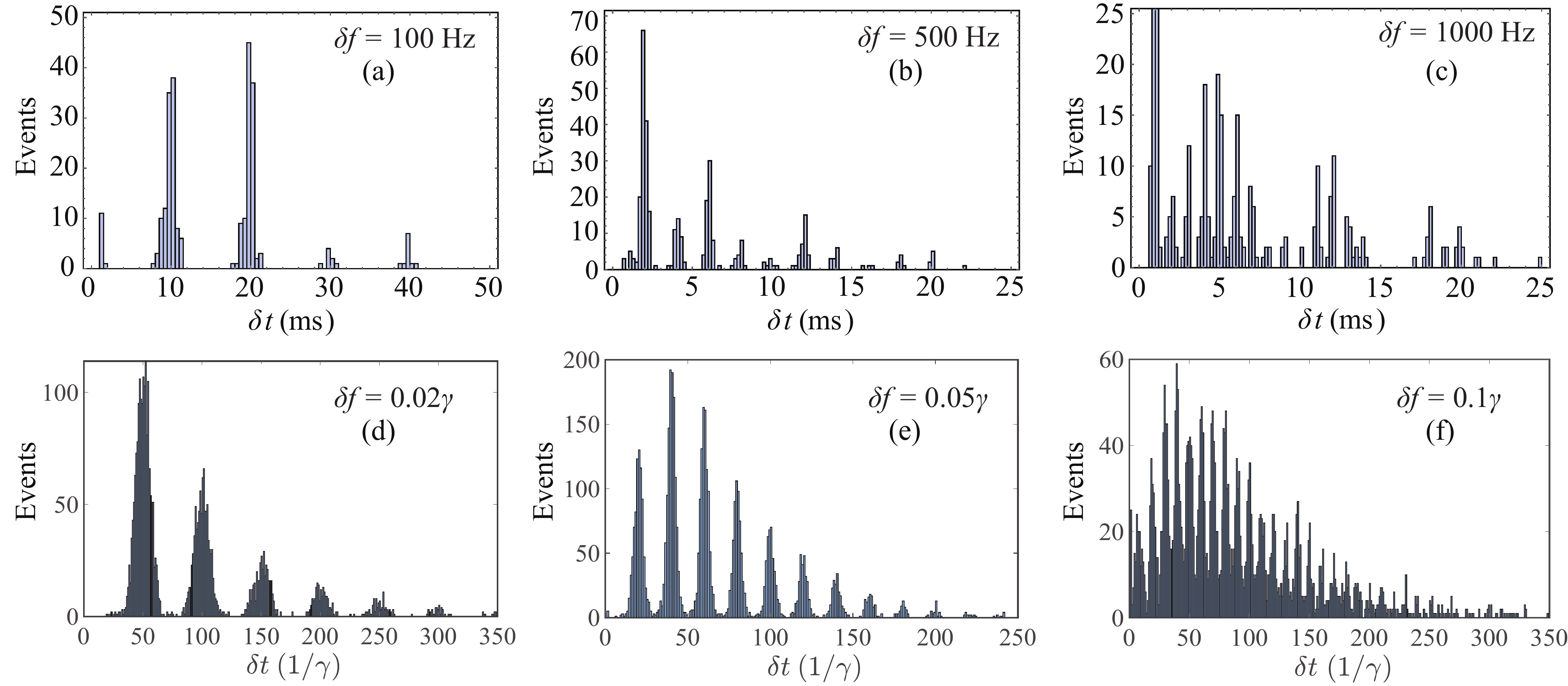}\\
\caption{\textbf{Experimental and theoretical distribution of the collective quantum jump times as a function of the MW frequency difference $\delta f$.} (a-c) Experimental distribution of the jump times for different relative detunings of the double-tune MW, $\delta f=$ 100 Hz (a), 500 Hz (b), and 1000 Hz (c). The distance between neighbouring peaks changes with $\delta f$. (d-f) Distribution of the jump times calculated through our theoretical model. The parameters are:  (d) $\delta f = 0.02\gamma$ and $\Delta=-3.9\gamma$, (e) $\delta f=0.05\gamma$ and $\Delta = -3.8\gamma$, (f) $\delta f=0.1\gamma$ and $\Delta = -3.7\gamma$. In all panels $\beta=-10$ dB. Other parameters are the same as those in Fig.~3(d-g).}\label{frequency}
\label{fig_s1}
\end{figure*}

Furthermore, Fig.~\ref{fig_s1} shows how  the distribution of $\delta t$ changes with the MW frequency difference $\delta f$, in both (a-c) experiment and (d-f) theoretical model. This illustrates that the sub-harmonic persists for varying $\delta f$.

\section{Simplified two-level atom approach}

In this section we study a simplified theoretical model, in which only two states per atom are considered. This simplification allows for more analytical insights on the observed phenomena. The aim is to  provide an overall qualitative understanding on the physics described by the previous more complex model.

\subsection{The model}

The base model we consider is a system of $N$ two-level atoms in which we have a ground state $|g\rangle$ and a Rydberg state $|r\rangle$. The Hamiltonian reads ($\hbar=1$):
\begin{equation}
H(t)=-\Delta(t)\sum_{j=1}^N |r\rangle \langle r|_j+\frac{\Omega}{2}\sum_{j=1}^N  (|r\rangle\langle g|_j+|g\rangle\langle r|_j) +\frac{V}{N-1}\sum_{j<k} |r\rangle \langle r|_j\otimes|r\rangle \langle r|_k,
\end{equation}
with a time dependent detuning:
\begin{equation}
\Delta(t)=\Delta-A\cos(2\pi \delta f t)\, .
\end{equation}
The amplitude $A$ and the frequency difference $\delta f$ define the periodic modulation. The atoms are driven with Rabi frequency $\Omega$. When found in the Rydberg state, they experience an all-to-all interaction with strength $V$.
\newline
The form of the periodic modulation of the detuning corresponds to modulating the energy of $|r\rangle$. For a slow modulation and small $A$, the $\Delta(t)$ could be implemented by considering $|r\rangle$ to be a dressed state, and $\delta f$ the frequency of a (slow and small) modulation of the Rabi frequency of the dressing laser (similar to the previous model).
We additionally consider spontaneous decay from $|r\rangle$ to $|g\rangle$ with a rate $\gamma$, and dephasing with a rate $\gamma_\mathrm{D}$. This leads to a master equation with two jump operators per atom: $L_{1,j}=\sqrt{\gamma}|g\rangle \langle r|_j$ and $L_{2,j}=\sqrt{\gamma_\mathrm{D}}|r\rangle \langle r|_j$.

\subsection{Mean-field dynamics}

For large system sizes, we can understand the dynamics of the above model through a mean-field approximation. The system is then described by the following non-linear equations of motion:
\begin{equation}
\dot{n}=\Omega \text{Im}(q)-\gamma n,
\end{equation}
\begin{equation}
\dot{q}=-i(\Delta(t)-nV)q-i\Omega n-\Gamma q+i\frac{\Omega}{2},
\end{equation}
where $n$ is the mean-field density of atoms in the Rydberg state, and $q$ is a complex number describing the coherence $ |r\rangle\langle g|$ in the mean-field limit. We have furthermore defined $\Gamma=(\gamma+\gamma_\mathrm{D})/2$.

\begin{figure*}[t!]
 \centering
 \includegraphics[width=1.0\columnwidth]{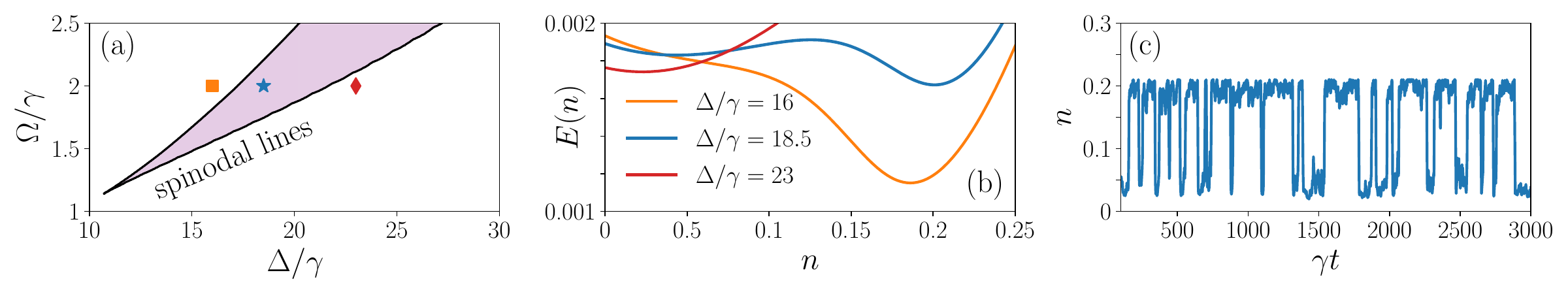}
 \caption{{\bf Bistability in the mean-field dynamics.} (a) Phase diagram of the mean-field model: in the white regions there is only one stable solution, in the purple region there are two. The other parameters are fixed to $V=100\gamma$, $\gamma_\mathrm{D}=10\gamma$. The orange square, blue star and red diamond correspond to the detunings $\Delta/\gamma=(16,18.5,23)$ respectively and $\Omega/\gamma=2$. (b) Effective potential $E(n)$ for the color points in (a). (c) Stochastic dynamics of the combined Eqs. (\ref{adiabatic_dynamics}) and (\ref{stochastic_detuning}) for $\Delta/\gamma=18.5$, $\kappa=0.1\gamma$ and $D=\gamma$.}
 \label{fig_bistability}
\end{figure*}

{\it Time-independent detuning. --} We first consider the case of constant detuning, i.e., $A=0$. In Fig.~\ref{fig_bistability}(a) we show a section of the phase diagram of the system. We see that there are two possible regions: one in which there is only one stable phase (white region) and one in which there is bistability (purple region). The two bistable phases correspond respectively to a phase with low Rydberg excitation and a phase with high one. Both regions are separated by spinodal lines. These lines merge at a critical point below which there is no phase transition.

We are interested in the case in which dephasing is strong and the dynamics of the coherences can be adiabatically eliminated. This leads to a single equation for the density of atoms in the Rydberg state:
\begin{equation}\label{adiabatic_dynamics}
\dot{n}=-\frac{\Omega^2\Gamma\big(n-\frac{1}{2}\big)}{\Gamma^2+(\Delta-nV)^2}-\gamma n.
\end{equation}
This dynamics can be understood in terms of an  effective potential energy for $n$, i.e. $\dot{n}=-\partial_n E(n)$. The latter is given by:
\begin{equation}
E(n)=\frac{\gamma n^2}{2}-\frac{\Omega^2}{V}\bigg(\frac{\Delta}{V}-\frac{1}{2} \bigg)\tan^{-1}\bigg(\frac{\Delta-nV}{\Gamma}\bigg)+\frac{\Omega^2 \Gamma}{2V^2}\text{log}\bigg(1+ \frac{(\Delta-nV)^2}{\Gamma^2}\bigg).
\end{equation}
The effective potential clearly illustrates the emergence of bistability in the purple region. This is shown in Fig.~\ref{fig_bistability}(b), in which we display  three potential landscapes  corresponding to the points marked in panel (a). For $\Delta/\gamma=16$ (orange square), the potential displays only one minimum, and the system settles in the high Rydberg excitation phase. Increasing the detuning to  $\Delta/\gamma=18.5$ (blue star), the system displays bistability, and the potential has two minima, one corresponding to the highly excited phase and one to the low excited one. Finally, if we increase further the detuning to $\Delta/\gamma=23$ (red diamond), only the low excited phase is stable and the potential displays only one minimum at small $n$.

\begin{figure*}[t!]
 \centering
 \includegraphics[width=1\columnwidth]{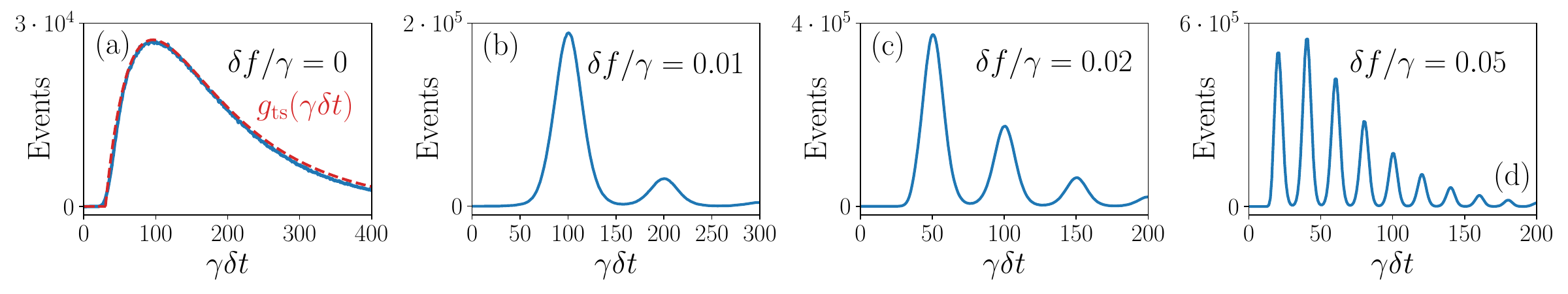}
 \caption{{\bf Histograms of difference of times $\delta t$.} The histograms are obtained from the stochastic dynamics of Eqs. (\ref{adiabatic_dynamics}) and (\ref{stochastic_detuning}) for $\Delta/\gamma=18.5$, $V=100\gamma$, $\gamma_\mathrm{D}=10\gamma$, $\Omega/\gamma=2$, $\kappa=0.1\gamma$ and $D=\gamma$. In (a) we consider $A=0$. The histogram is fitted by the function $g_\mathrm{ts}(\gamma \delta t)$ with parameters $C=5.68\cdot 10^6$, $\gamma_1/\gamma=0.025$, $\gamma_2/\gamma=0.0083$, $\gamma \delta t_0=15$. (b) $A=3\gamma$ and $\delta f/\gamma=0.01$. (c) $A=3\gamma$ and $\delta f/\gamma=0.02$. (d) $A=3\gamma$ and $\delta f/\gamma=0.05$.}
 \label{fig_counts}
\end{figure*}

{\it Stochastic dynamics. --} Besides the mean-field approximation, we  include noise and fluctuations by considering a stochastic process for the detuning. Then, we make the substitution $\Delta\to \Delta+\Delta_\mathrm{S}$ in Eq. (\ref{adiabatic_dynamics}), where  $\Delta_\mathrm{S}$ follows an Ito stochastic differential equation:
\begin{equation}\label{stochastic_detuning}
d\Delta_\mathrm{S}=-\kappa \Delta_\mathrm{S} dt+\gamma\sqrt{D}dB_t,
\end{equation}
where $dB_t$ is the increment of a Wiener process, $\kappa$ ensures $\Delta_\mathrm{S}$ fluctuates around zero and $D$ is the strength of the noise. In the bistable region, the combined dynamics of Eq.~(\ref{adiabatic_dynamics}) and Eq.~(\ref{stochastic_detuning}) leads to the system randomly switching between the phases with high and low Rydberg excitation, see Fig.~\ref{fig_bistability}(c). Moreover, in Fig.~\ref{fig_counts}(a) we show  an histogram of the times between consecutive upward jumps, i.e. $\delta t$.

The  histogram of Fig.~\ref{fig_counts}(a) can be fitted by the function:
\begin{equation}
g_\mathrm{ts}(\delta t)=C\frac{\gamma_1 \gamma_2}{\gamma_1-\gamma_2}\big[e^{-\gamma_2(\delta t-2\delta t_0)}-e^{-\gamma_1(\delta t-2\delta t_0)} \big]H(\delta t-2\delta t_0),
\end{equation}
where $C$ is a fitting constant and $H(x)$ is a Heaviside step function. This function corresponds to the (rescaled by $C$) probability density function for the time intervals between two consecutive upward jumps, $\delta t$, for a two state classical stochastic process with a latency time $\delta t_0$. In this two state stochastic process, the probability density for jumping out of each state after a time $t$ is given by:
\begin{equation}
g_\mathrm{ts,1}(t)=\gamma_1 e^{-\gamma_1 (t-\delta t_0)}H(t-\delta t_0), \quad g_\mathrm{ts,2}(t)=\gamma_2 e^{-\gamma_2 (t-\delta t_0)}H(t-\delta t_0),
\end{equation}
where $\gamma_{1,2}$ models the rate of jump out of each phase and $\delta t_0$ is a latency time. In this model, the two states describe the two bistable phases, while the latency time models faster relaxation process within each phase that prevent jumps to occur in timescales shorter than $\delta t_0$. Finally notice that for large $\delta t$, $g_\mathrm{ts}(\delta t)$ displays an exponential decay with $\delta t$, similarly to what is observed experimentally.

{\it Stochastic dynamics with time-modulated detuning. --} Finally we consider the dynamics of Eqs. (\ref{adiabatic_dynamics}) and (\ref{stochastic_detuning}) but with a time-modulated detuning $A\neq0$. We consider the parameters corresponding to the system in the bistable regime [blue star of Fig.~\ref{fig_bistability}(a)]. We analyze the effects of the time-modulation in the histogram of $\delta t$. The results for different frequencies $\delta f$ are shown in Fig.~\ref{fig_counts}(b) to (d). We observe that the histogram develops a structure of peaks, whose separation coincides with the period of modulation. The larger the frequency modulation, the more higher-order peaks are activated. This is analogous to what has been observed experimentally and modeled with the more complex theoretical model.

